\documentclass[a4paper,11pt]{book}
\usepackage{graphicx}
\usepackage[numbers,compress]{natbib}
\usepackage{bm}
\usepackage{amsmath,amssymb}

\def\beq{\begin{eqnarray}}
\def\eeq{\end{eqnarray}}

\def\be{\begin{equation}}
\def\ee{\end{equation}}
\def\beq{\begin{eqnarray}}
\def\eeq{\end{eqnarray}}

\def\ii{{\rm i}}

\def\IL{\relax{\rm I\kern-.18em L}}

\def\f{\frac}

\title{{\bf\huge Rotating analogue black holes:
Quasinormal modes and tails, 
superresonance, and sonic bombs and plants in the 
draining bathtub acoustic hole}\\
\centerline{}
\centerline{\bf Jos\'{e} P. S. Lemos}
{\small\it Centro Multidisciplinar de Astrof\'{\i}sica - CENTRA,
Departamento de F\'{\i}sica, Instituto Superior T\'ecnico - IST,
Universidade de Lisboa - UL, \hfill\newline Av. Rovisco Pais 1,
1049-001 Lisboa, Portugal\\}
\centerline{}
\centerline{}
\small In {\it Analogue spacetimes: The first thirty years}, 
eds.~L.~C.~Crispino et
al (Editora Livraria da F\ii sica, S\~ao Paulo 2013), p. 145;
Proceedings
of the II Amazonian Symposium on Physics ``Analogue Models of
Gravity: 30 Years Celebration of the publication of Bill Unruh's paper
``Experimental Black-Hole Evaporation?"". 
Based on the invited lecture at the Symposium realized in the
Universidade Federal do Par\'a, Bel\'em do Par\'a, Brazil, June
1$\,$-$\,$3, 2011.  }
\date{}
\begin{document}

\maketitle

\chapter*{Rotating analogue black holes:
Quasinormal modes and tails, 
superresonance, and sonic bombs and plants in the 
draining bathtub acoustic hole}
%Rotating acoustic holes: Quasinormal modes and tails, 
%superresonance, and sonic bombs and plants in the draining bathtub

\author{Jos\'{e} P. S. Lemos\footnote{joselemos@ist.utl.pt}\\
\textit{Centro Multidisciplinar de Astrof\'{\i}sica - CENTRA,
Departamento de F\'{\i}sica, Instituto Superior T\'ecnico - IST,
Universidade de Lisboa - UL, Av. Rovisco Pais 1, 1049-001
Lisboa, Portugal}\\
%\and
%Author\footnote{author email}\\
%\textit{Affiliation}\\
}

\textbf{Abstract:} The analogy between sound wave propagation and
light waves led to the study of acoustic holes, the acoustic analogues
of black holes. Many black hole features have their counterparts in
acoustic holes. The Kerr metric, the rotating metric for black holes
in general relativity, has as analogue the draining bathtub metric, a
metric for a rotating acoustic hole. Here we report on the progress
that has been made in the understanding of features, such as
quasinormal modes and tails, superresonance, and instabilities when
the hole is surrounded by a reflected mirror, in the draining bathtub
metric. Given then the right settings one can build up from these
instabilities an apparatus that stores energy in the form of amplified
sound waves.  This can be put to wicked purposes as in a bomb, or to
good profit as in a sonic plant.

\newpage
\section*{1. Introduction}
\subsection*{1.1 Black holes}

Black holes, the most fascinating prediction of general relativity,
are the simplest objects which can be built out of spacetime itself
\cite{mtw}. They are hard to detect. Astrophysical black holes are
very far away and electromagnetically can only be inferred
indirectly. The manufacturing of TeV scale black holes is a dream
(see for a review \cite{tev}),
whereas Planckian scale black holes are far beyond current and future
energetic technological capabilities \cite{lemosreview1996}.

Understanding classical physical phenomena that occurs with black
holes is important. For instance, black hole quasinormal modes (QNMs),
which appear when a black hole is naturally perturbed, can test black
hole stability and should provide means to identifying black hole
parameters like the mass $M$, the angular momentum $J$, and the charge
$Q$, from the real $\omega_R$ and imaginary $\omega_I$ parts of the
mode frequencies
\cite{qnms1vish,qnms2press,SchutzWillApJ1985,WillIyerPRD1987,NollertPRD1993,MotlATMP2003,MotlNeitzkeATMP2003,BertiKokkotasPRD2003,CardosoLemosYoshida2PRD2004,KonoplyaPRD2003}.
Power-law tails that rise at the end of the perturbation
also give insight into the black hole features and parameters
\cite{PricePRD1972}.  Moreover, the Penrose process
\cite{PenroseNC1969}, superradiance
\cite{ZeldovichJETPLett1971,ZeldovichJETP1972,PTAPJ1974,UnruhPRD1974,bekensuperr},
and black hole instabilities, such as the black hole bomb instability
\cite{PressTeukolskyNature1972,CDLYPRD2004}, as well as no-hair
theorems and their possible tests \cite{EcheverriaPRD1989,nohairtests}, 
are phenomena that help in the making of the whole picture of what a 
black hole is.

Understanding of quantum phenomena is also important.  Hawking
show\-ed that when quantum effects are taken into account, black holes
emit thermal radiation \cite{hawkingrad}. This led to a number of
fundamental questions, such as the information puzzle, black hole
entropy (which in turn, following a conjecture put forward in
\cite{HodPRL1998}, could be related to quasinormal modes with a large
imaginary part and give direct information on the degrees of freedom
of the quantum area cells of the black hole horizon, see also
\cite{MotlATMP2003,MotlNeitzkeATMP2003,BertiKokkotasPRD2003,CardosoLemosYoshida2PRD2004,KonoplyaPRD2003}),
and the black hole final state (for a review see \cite{lemosreview}).

\subsection*{1.2 Acoustic holes}

There is the question posed by many physics students of why the
velocity of sound is different from the velocity of light. There are
many differences, of course, but in a way Unruh \cite{unruhanalogue}
brought them a bit closer to each other. 
Indeed, prospects for having a black
hole in the hand changed when in 1981 Unruh \cite{unruhanalogue}
realized that for a fluid moving with a space-dependent velocity
which exceeds the sound velocity of the medium one gets the equivalent
of an event horizon for sound waves. The concept of a dumb hole or
acoustic hole, the acoustic analogue of a black hole, was created. 
This concept has been further developed, see e.g., 
\cite{VisserCQG1998,SchutzholdUnruhPRD2002,VisserWeinfurtnerCQG2004}, 
and  \cite{reviewvisser} for a review, where, 
among other things, similarities and differences between 
the rotating draining bathtub acoustic hole
and the Kerr black hole are displayed. 
Many other different kinds of analogue holes have been devised involving
not only fluids, but also condensed matter systems, slow light 
devices, and other models
\cite{RezPRD2000,GarayAngCirZolPRA2001,BarLibVisCQG2001}.
Within these models, phonon radiation, the sonic analogue of 
Hawking radiation and one of the most important effects in 
these physical systems,  has been studied 
with some care 
\cite{UnruhPRD1995,VisserPRL1998,FischerVolovikIJMPD2001,Unruhetal2010}.

There are certainly other phenomena that occurs with true black holes
that also occurs with acoustic holes. 
For instance, there is analogous
geodesic and causal structure, and Carter-Penrose diagrams can be drawn
\cite{BarceloLiberatiSonegoVisserNewJPhys2004}.
Quasinormal modes and power-law tails are also expected in acoustic models
\cite{BertiCardosoLemosPRD2004,CardosoLemosYoshidaPRD2004}, see also
\cite{DolanOliveiraCrispino2011,MarquesMScThesis2011}.
As well, in analogy with general relativistic effects in black holes,
we also expect in acoustic holes the phenomenon of superresonance
\cite{BasakMajumdarCQG20031,BasakMajumdarCQG20032}
(see also \cite{SchutzholdUnruhPRD2002}),
which in some cases can be amplified into a sonic bomb 
\cite{BertiCardosoLemosPRD2004}, or depending on
the use into a sonic plant.

\subsection*{1.3 Acoustic holes in this presentation}

We use the rotating draining bathtub acoustic hole
\cite{VisserCQG1998,SchutzholdUnruhPRD2002,VisserWeinfurtnerCQG2004}, 
which is an analogue of
the Kerr black hole.  We study the QNMs of this rotating draining
bathtub acoustic hole
\cite{BertiCardosoLemosPRD2004,CardosoLemosYoshidaPRD2004}, see also
\cite{DolanOliveiraCrispino2011,MarquesMScThesis2011}, and the
power-law tails that appear long after the initial perturbation
\cite{BertiCardosoLemosPRD2004}. Then, we compute the reflection
coefficients for superresonant scattering in this acoustic hole
background
\cite{BertiCardosoLemosPRD2004,CardosoLemosYoshidaPRD2004}. Enclosing
the acoustic hole by a reflecting mirror we install an instability
into the system thereby turning it into a sonic bomb  
\cite{BertiCardosoLemosPRD2004} or a sonic power
plant. The worked reported in this
review has been done in collaboration with Emanuele Berti, Vitor
Cardoso, and Shijun Yoshida
\cite{BertiCardosoLemosPRD2004,CardosoLemosYoshidaPRD2004}.

\section*{2.
The draining bathtub acoustic hole: the geometry}\label{bathtub}

The draining bathtub model first introduced by Visser 
\cite{VisserCQG1998}
for a rotating acoustic hole is the starting point
(see also \cite{SchutzholdUnruhPRD2002,VisserWeinfurtnerCQG2004}).
The fluid moves in a plane, it is a 2-dimensional flow, which can be made
3-dimensional by the addition of another trivial dimension perpendicular
to the original plane.

Consider a fluid having (background) density $\rho$.  Assume the fluid
to be locally irrotational (vorticity free), barotropic and
inviscid. Use polar coordinates $(r,\phi)$. 
From the equation of continuity, the radial component of the
fluid velocity satisfies $\rho v^r\sim 1/r$. Irrotationality implies
that the tangential component of the velocity satisfies $v^\phi\sim
1/r$. By conservation of angular momentum we have $\rho v^\phi\sim
1/r$, so that the background density of the fluid $\rho$ is
constant. In turn, this means that the background pressure $p$ and the
speed of sound $c$ are constants. Thus one can put
\begin{equation}
{\vec v}=\frac{-A\,\, {\hat {\rm e}_r}+
B \,\, {\hat {\rm e}_\phi}}{r}\,,
\end{equation}
where $A$ and $B$ are constants and ${\hat {\rm e}_r}$ and 
${\hat {\rm e}_\phi}$ are the unit vectors in 
the radial and azimuthal directions, respectively.
This flow velocity can be obtained as the gradient of a velocity
potential, $\vec v=\nabla \bar{\psi}$, where
\begin{equation}\label{vpot}
\bar \psi=A\log r+B\,\phi\,.
\end{equation}

Perturb the fluid, and encode the perturbation in the velocity potential,
so that the new velocity potential is $\bar{\psi}+\psi$, $\psi$ being
the perturbation. Unruh \cite{unruhanalogue} 
first realized that the propagation of a sound wave
in a barotropic inviscid fluid with irrotational flow is described by
the following equation for $\psi$, 
\begin{equation}
\nabla_{\mu}\nabla^{\mu}\psi=0\,,
\label{kgeq}
\end{equation}
where the operator $\nabla_{\mu}$ denotes
covariant derivative, and $\mu$ runs from 0 to 2, 
with 0 denoting a time coordinate and 1,2 spatial coordinates.
This is the Klein-Gordon equation for a
massless field, in this case the perturbed velocity potential $\psi$, 
in a Lorentzian acoustic curved geometry. 

In our case, the acoustic metric describing the
propagation of sound waves in this draining bathtub 
fluid flow is $(2+1)$-dimensional and given by
\cite{VisserCQG1998}, 
\begin{equation}
ds^2=
-\left (c^2-\frac{A^2+B^2}{r^2} \right )dt^2
+\frac{2A}{r}drdt-2Bd\phi dt+dr^2+
r^2d\phi^2\,,
\label{metric1}
\end{equation}
where $t$ is the time coordinate.
In the non-rotating limit $B=0$ the metric (\ref{metric1}) reduces to
a standard Painlev\'e-Gullstrand-Lema\^itre type metric. If one adds
a $dz^2$ term the metric turns into a $(3+1)$-dimensional metric.
The acoustic horizon is located at 
\begin{equation}
r_H=\frac{A}{c}\,. 
\label{rh}
\end{equation}
The ergosphere forms at
\begin{equation}
r_{es}=\frac{(A\,^2+B\,^2)^{1/2}}{c}\,, 
\label{res}
\end{equation}
and the angular velocity of the horizon is 
\begin{equation}
\Omega=\frac{B}{r_H^2}\,. 
\label{ome}
\end{equation}

\begin{figure}
[t]
\centerline{\mbox{
\includegraphics[height=5cm,angle=0]{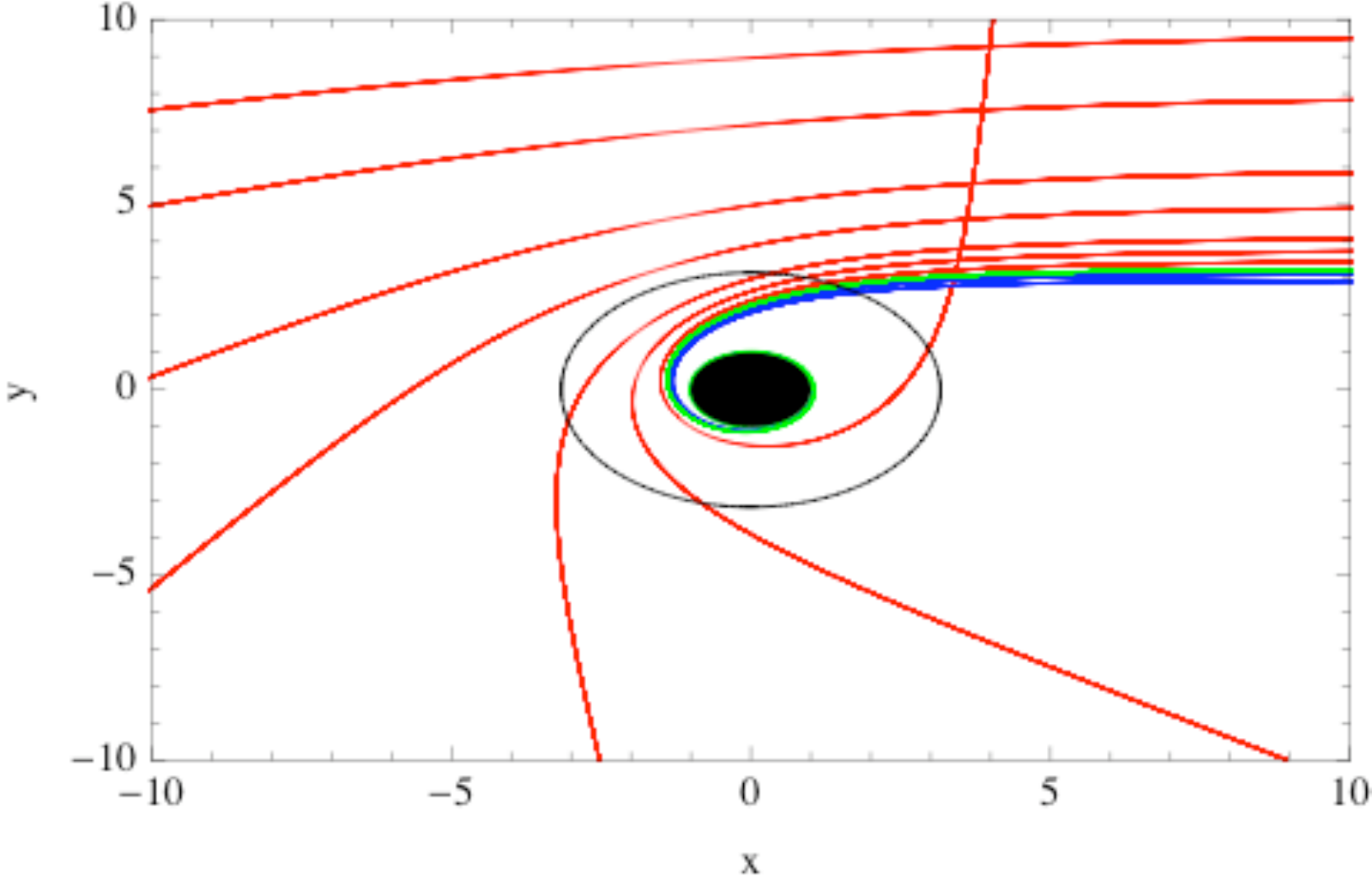}
}}
\centerline{\mbox{
\includegraphics[height=5cm,angle=0]{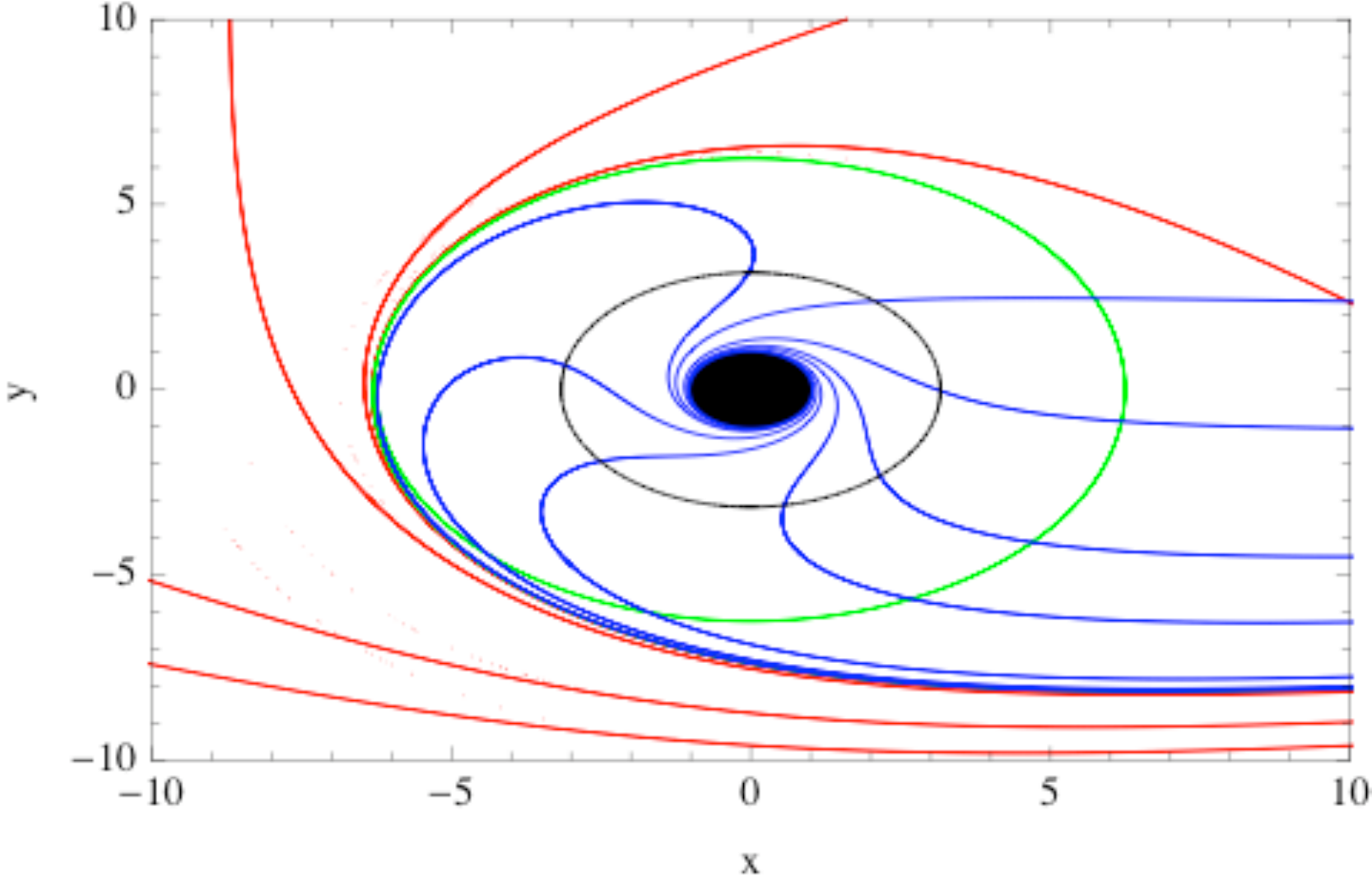}
}}
\caption{\small Rotating and counterrotating geodesics in the 
draining bathtub acoustic hole.}
\label{fig:geodesics12}
\end{figure}

It is useful sometimes 
to see the metric in a Boyer-Lindquist Kerr metric form. Transforming
then to Boyer-Lindquist
coordinates $(\widetilde{t},r,\widetilde{\phi})$, one finds
\begin{equation}
ds^2=
-\left(1-\frac{A^2+B^2}{c^2r^2} \right)c^2 d\widetilde{t}^2+
\left(1-\frac{A^2}{c^2r^2} \right )^{-1}dr^2
-2B d\widetilde{\phi}d\widetilde{t}+r^2d\widetilde{\phi}^2\,.
\label{kerr}
\end{equation}
The geodesics of a test particle in  Boyer-Lindquist coordinates 
can be seen in  Fig.~\ref{fig:geodesics12} 
\cite{MarquesMScThesis2011}.

\section*{3. The basic equations for the scalar field perturbation 
$\mathbf\psi$ in 
the draining bathtub acoustic hole}\label{basiceqs}

Now we want to solve 
the Klein-Gordon equation (\ref{kgeq}) for the
massless field $\psi$
in the Lorentzian acoustic geometry, which in 
our case takes the form  (\ref{metric1}). We want to determine the
modes of the field $\psi$.
We can separate variables by the substitution
\begin{equation}
\psi(t,r,\phi)=R(r)e^{-\ii (\omega t-m\phi)}\,,
\end{equation}
where $R(r)$  is a function of the coordinate $r$,
$\omega$ is the frequency of the mode and $m$ its 
azimuthal wave number.
We now introduce a tortoise coordinate $r_*$
defined by the condition
${dr_*}/{dr}= {1}/({1-A^2/c^2r^2})$.
Explicitly,
$r_*=r+\f{A}{2c}\log\left|\f{cr-A}{cr+A}\right|$, 
which in turn can be inverted to give $r(r_*)$. 
To obtain a Schr\"odinger-like equation we set
$R(r)=Z(r) H(r)$ and find, first,
$Z$ as a function of $(r,A,B,m)$, 
$Z(r)=r^{1/2}\,f(r,A,B,m)$ which is not necessary
to give here explicitly. Then 
the Schr\"odinger-like equation for $H(r)$ is then
\begin{equation}
H_{,r_* r_*}+
\left\{\f{1}{c^2}\left(\omega-\f{Bm}{r^2}\right)^2-
\left(\f{c^2r^2-A^2}{c^2r^2}\right)
\left[\f{1}{r^2}\left(m^2-\f{1}{4}\right)+
\f{5A^2}{4r^4 c^2}\right]\right\}
H=0\,.
\label{waveequation}
\end{equation}
The wave equation (\ref{waveequation}) can be cast in a more
useful form by performing the following rescalings, $\widehat r=rA/c$,
$\widehat{\omega}=\omega A/c^2$, $\widehat{B}=B/A$. We then get
the Schr\"odinger-like equation 
\begin{equation}
H_{,\widehat r_* \widehat r_*}+Q H=0\,,
\label{waveequation2}
\end{equation}
with the generalized potential given by, 
\begin{equation}
Q\equiv 
\left\{ \left(\widehat{\omega}-\f{\widehat{B}m}{\widehat r^2}\right)^2-
V \right\}\,,
\quad
V\equiv \left(\f{\widehat r^2-1}{\widehat r^2}\right)
\left[\f{1}{\widehat r^2}
\left(m^2-\f{1}{4}\right)+\f{5}{4\widehat r^4}\right]\,.
\label{Qdef}
\end{equation}
The rescaling is equivalent to set $A=c=1$ in 
the master wave equation,
and at the same time 
the acoustic horizon obeys $\widehat r_H=1$. 
In addition, the horizon angular velocity is $\Omega=B/r_H^2$, i.e., 
$\Omega=\widehat B$ in these units.
Unless otherwise stated, we will
omit hats in all quantities. The
rescaled wave equation (\ref{waveequation2}) will be the starting
point of our analysis of QNMs 
\cite{BertiCardosoLemosPRD2004,CardosoLemosYoshidaPRD2004}
(see also \cite{DolanOliveiraCrispino2011}).
Then we study quasinormal modes 
\cite{BertiCardosoLemosPRD2004,CardosoLemosYoshidaPRD2004},
power-law tails \cite{BertiCardosoLemosPRD2004}, 
superresonant
phenomena \cite{BertiCardosoLemosPRD2004}, and sonic bombs 
\cite{BertiCardosoLemosPRD2004} and plants.

\section*{4. Evolution of the perturbations 
in the draining bathtub acoustic hole: Quasinormal modes and tails}

\subsection*{4.1 Generic features}

The evolution of the perturbations of an acoustic hole 
has the same pattern as
the evolution of perturbed black holes in
spacetime. It can be divided into three parts:
(i) Initially, there is the prompt response, which 
is dependent on the initial conditions.
(ii) At an intermediate stage, the signal is 
predominantly an oscillating 
exponential which 
decays, giving rise to 
the quasinormal mode (QNM) oscillations,
dependent  on the
black hole parameters alone, namely, mass, and angular momentum.
(iii) At late times, the backscattering off the curvature
creates a tail, a power law falloff of
the field, which has a strong 
dependence on the asymptotic far region.
Let us see in detail what really happens in the rotating 
draining bathtub acoustic hole.

\subsection*{4.2 Boundary conditions and asymptotic solutions}

The QNMs of the rotating acoustic hole can be setup by defining
appropriate boundary conditions and solving the corresponding
eigenvalue problem, Eqs.~(\ref{waveequation2})-(\ref{Qdef}).  The
boundary conditions are that close to the event horizon the solutions
behave as $H \sim e^{- \ii\left (\omega-Bm\right)r_* }$, i.e., there
are only ingoing waves.  At spatial infinity the solutions of
(\ref{waveequation2})-(\ref{Qdef})
behave as $H \sim e^{+ \ii\omega r_*}$, i.e.,
there are only outgoing waves.

Then, for the wave equation (\ref{waveequation2})-(\ref{Qdef}) with
the boundary conditions as given above, and assigned values of $B$ and
$m$ (the rotational parameter and the angular index $m$, respectively)
there is a discrete and infinite set of QN frequencies, $\omega_{QN}$.

The QN frequencies are complex numbers. 
Since the time
dependence is given by $e^{-\ii \omega t}$, 
the imaginary part
describes the decay or growth of the perturbation. 
As it is expected the hole is
stable against small perturbations,
and thus $\omega_{QN}$
should have a negative imaginary part for 
an exponential time decay of the perturbation.

As it is usual, 
the QN frequencies $\omega_{QN}$ are ordered through
the absolute value of their imaginary part, the modes being labeled by
an integer $n$. The fundamental mode $n=0$ will have the smallest
imaginary part, in modulus.

\subsection*{4.3 Slowly decaying modes of non-rotating holes} 

The ringing behavior of a classical
perturbation is controlled by the
lowest QNMs. 
The overtones with higher frequency
have a larger imaginary part, and so 
are damped faster and play no role.  
The fundamental mode with $n=0$ 
is effectively responsible for the response of the 
acoustic hole to
exterior perturbations.

The use of a WKB approximation with 
corrections up to the sixth order
is certainly good enough. Such an approximation was developed in 
\cite{SchutzWillApJ1985,WillIyerPRD1987}, 
and to use it, $Q$  in Eqs.~(\ref{waveequation2})-(\ref{Qdef})
has to have a 
single maximum (see also \cite{KonoplyaPRD2003}). 
For non-rotating holes, i.e., $B=0$,  this is true 
for $m\neq0$. 
\begin{table}
\centering
\begin{tabular}{@{}c|c|c|c@{}}  
\hline
\hline
$m$ &$\omega _{QN}^{(1)}$  &$\omega _{QN}^{(3)}$   &$\omega _{QN}^{(6)}$\\
\hline
\hline
%0  &  & 0.035-0.609i &              \\
1  & 0.696-0.353i & 0.321-0.389i &0.427-0.330i\\
2  & 1.105-0.349i & 0.940-0.353i &0.945-0.344i\\
3  & 1.571-0.351i & 1.465-0.353i &1.468-0.352i \\
4  & 2.054-0.352i & 1.975-0.353i &1.976-0.353i \\
\hline 
\hline
\end{tabular}
\caption{\small The fundamental mode $n=0$ of the rotating 
draining bathtub acoustic hole is shown 
for four different 
values of $m$. The first, third and sixth 
orders WKB approximation are displayed.}
\label{ltable}
\end{table}
In Table 1 we present the fundamental mode $n=0$, for different 
values of $m$ in the WKB approximation in several orders.
The real part of $\omega _{QN}$ 
scales nearly with $m$ and the imaginary part is
approximately constant as a function of $m$.
In the large $m$ limit one
can show from the WKB formula \cite{SchutzWillApJ1985}
that
$\omega_{QN}$ behaves as
\begin{equation}
\omega \sim \frac{m}{2}-i\,\frac{2n+1}{2\sqrt{2}}\,,\,m 
\rightarrow \infty\,,
\label{largembehav}
\end{equation}  
and indeed already for $m=4$ formula 
(\ref{largembehav}) yields excellent concordance
with the results shown in Table \ref{ltable}. 

To have some idea of the orders of magnitude 
involved, consider an $m=2$
mode. Take typical values for 
a wave analogue experiment, such
as those that can be directly 
inferred from \cite{SchutzholdUnruhPRD2002},
namely, $c=0.31\,$m/s, $r_H=1\,$m, $A=cr_H=0.31\,$m$^2$/s, and 
obtain for
the fundamental $n=0$ QNM with $m=2$, a frequency $\omega_R=0.945
\times c^2/A=0.293$ Hz and a damping timescale $\tau=1/|\omega_I| =
1/0.344 \times A/c^2=9.37$ s.

For the case $m=0$ the situation is more complicated, 
the WKB approximation is not usable
because $Q$ has more than one maxima. Using other sophisticated
methods one can show that there are no $m=0$ QNMs
\cite{DolanOliveiraCrispino2011}.

\subsection*{4.4 Slowly decaying modes of rotating holes}

For rotating holes, the generalized potential, 
Eq.~(\ref{Qdef}), has more than one
extremum, so one has to use the WKB approximation with 
some care. 
It is necessary to take the only root that 
gives the correct non-rotating limit as $B\rightarrow 0$.  
It is this solution the one
that corresponds to the QNMs of the rotating hole. 

\begin{figure}
\centerline{\mbox{
\includegraphics[height=5cm]{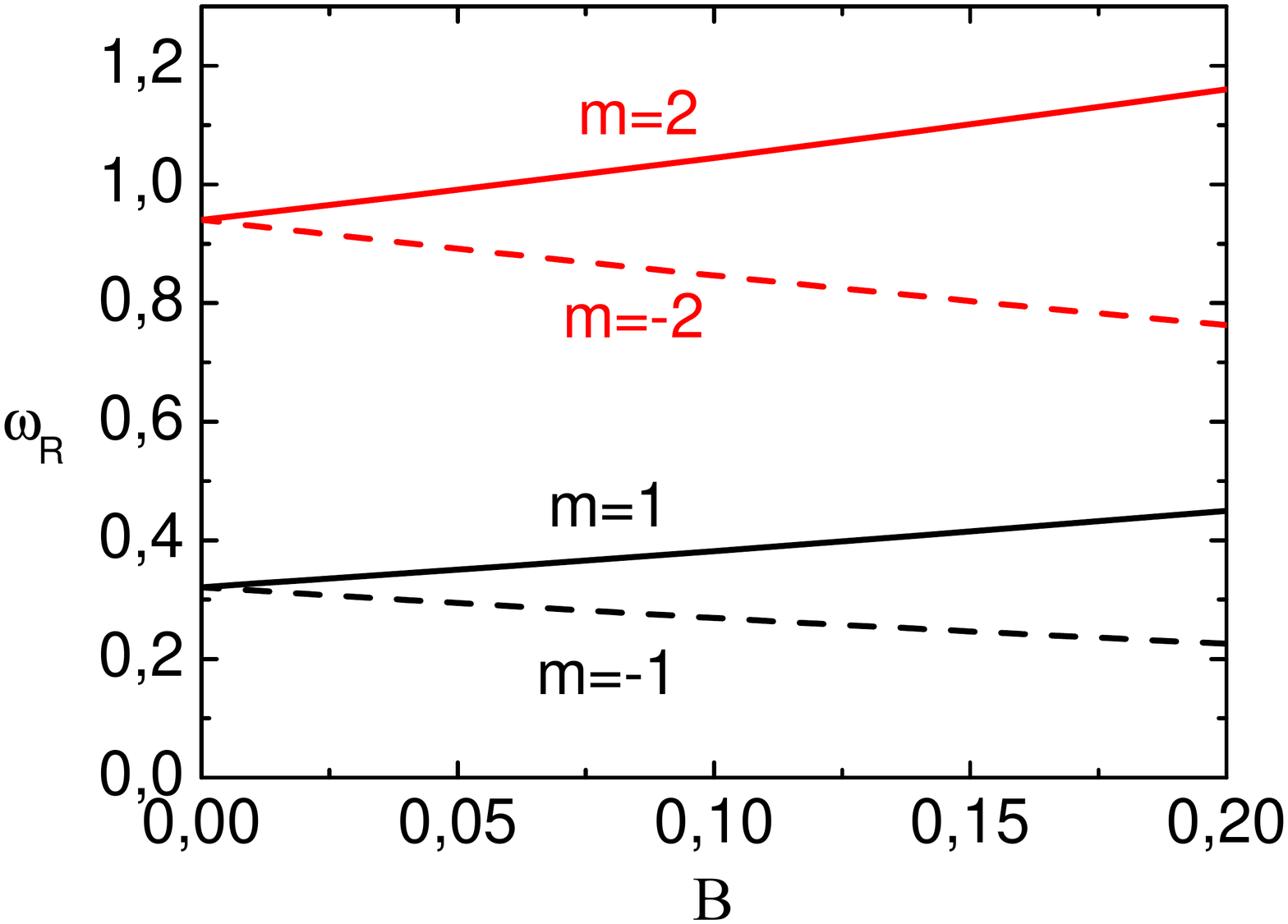}
\includegraphics[height=5cm]{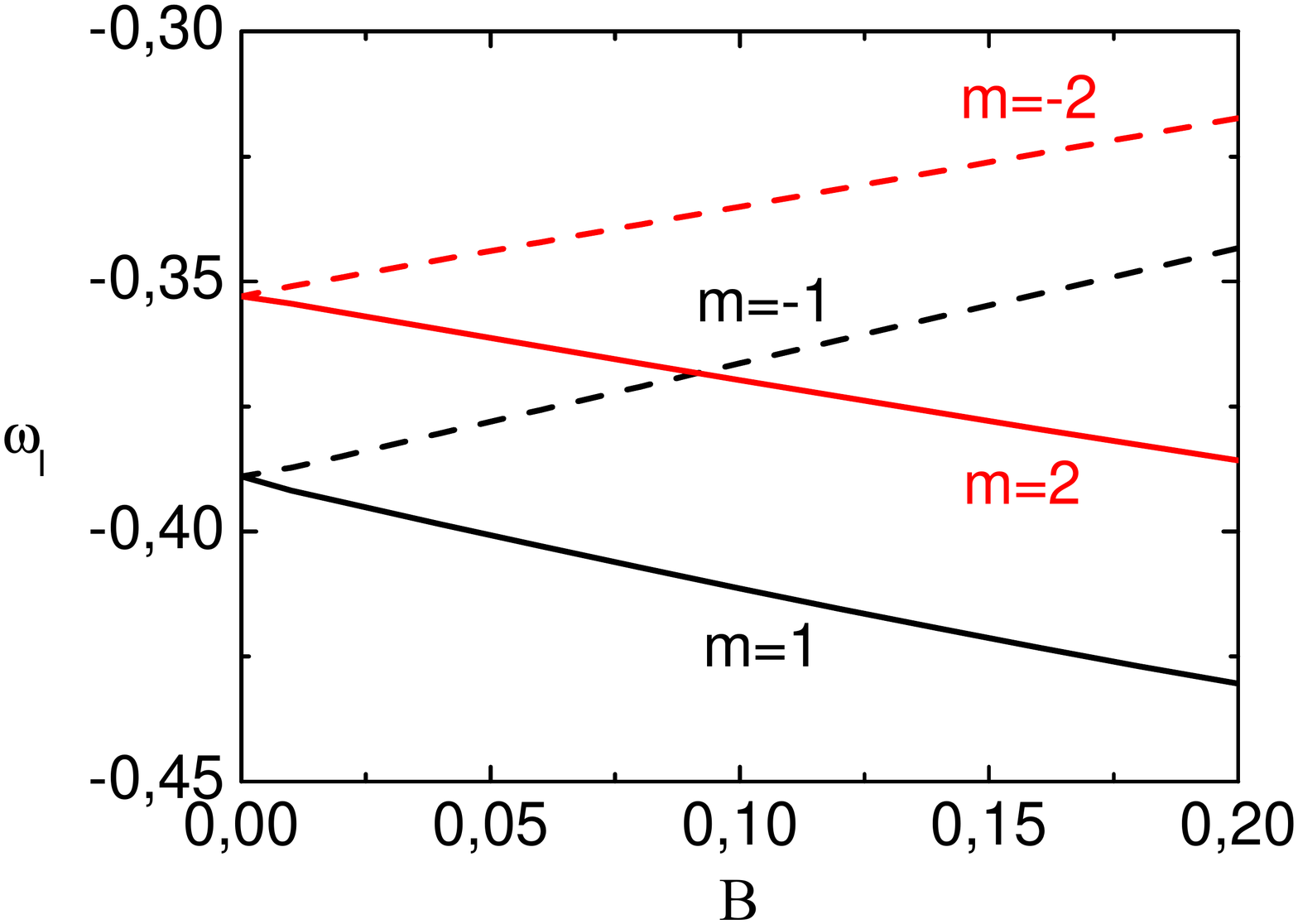}
}}
\caption{\small The real and imaginary frequencies
of QNMs for the rotating 
draining bathtub acoustic hole. 
These frequencies, 
$\omega_R$ and $\omega_I$ respectively, 
are shown as a function 
of $B$ (the rotating
parameter) 
for different values of $m$ (the perturbation azimuthal
number).}
\label{rotwkb}
\end{figure}

The WKB results we present can be trusted for $B\lesssim 0.1$.  In
Figure \ref{rotwkb} the QNM frequencies for the draining bathtub
acoustic hole are showed.  In the left plot the real part of the
fundamental QN frequency $\omega_R$ as a function of the rotation
parameter $B$, for selected values of $m$, is shown. In the right
plot the same is done for the imaginary part $\omega_I$.  Note that
$\omega_R$ and $|\omega_I|$ increase with rotation for $m>0$, while
$\omega_R$ and $|\omega_I|$ decrease with rotation for $m<0$.  Figure
\ref{rotwkb} seems to point out that it is possible an instability
sets in for large $Bm$. To settle this issue an analysis beyond the
WKB approximation is required.

Such an analysis of the QNMs in the high rotation regime, where the
WKB approximation breaks down, was performed in
\cite{CardosoLemosYoshidaPRD2004}, using a continued fraction analysis
and numerical techniques. This was supplemented in the thorough work
presented in \cite{DolanOliveiraCrispino2011} using a variety of
techniques.  It was found, as expected, that there are no
instabilities for large $Bm$.

An interesting point worth mentioning is that the change of the QN
frequency with rotation, although not drastic, can be used to apply a
fingerprint analysis of the acoustic hole parameters (see
\cite{EcheverriaPRD1989} for the black hole case). Once we measure at
least two QNM frequencies we may infer the hole parameters $A$ and
$B$.  Carrying out such experiments in the laboratory
can shed some light on
the applicability of similar ideas to test the no-hair theorem in the
astrophysical context \cite{nohairtests}.

\subsection*{4.5 Highly damped modes}  

QNMs with a large imaginary part, i.e., with a very large overtone
number $n$, have interest for black holes in general relativity and
other alternative theories of gravitation.
Hod \cite{HodPRL1998} proposed that
these modes could be related to black hole area quantization. 
An analytical calculation of highly damped QNMs was
first carried out by Motl for the Schwarzschild black hole
\cite{MotlATMP2003,MotlNeitzkeATMP2003}.
This analytical
results are in  agreement with numerical data 
\cite{BertiKokkotasPRD2003,CardosoLemosYoshida2PRD2004}
(see also
\cite{NollertPRD1993}). 

To start with, let us consider a non-rotating hole, $B=0$.
Afterward we add rotation. For $B=0$
the wave equation  reduces to
\begin{equation}
\frac{d^2H}{dr_*^2}+
\left [ \omega ^2-
V(r)
\right] H=0\,,
\end{equation}
with
\begin{equation}
V(r)\equiv \left ( 1-\frac{1}{r^2}\right ) 
\left (\frac{m^2-1/4}{r^2}+\frac{5}{4r^4} \right )\,.
\end{equation}
Following \cite{MotlATMP2003,MotlNeitzkeATMP2003} 
these modes are determined from $V(r)$ near the singular point
$r=0$. Thus 
$V \sim -{5}/{36r_*^2}$
(since $V \sim -{5}/{4r^6}$
and  $r_* \sim -r^3/3$). 
Write $V={(j^2-1)}/{4r_*^2}$ as one should 
for a field of spin $j$ \cite{MotlATMP2003,MotlNeitzkeATMP2003}.
Ten one finds that in our case
$j=2/3$. Thus the scalar field of the perturbation 
has spin of $2/3$. A weird spin. 
The result in \cite{MotlATMP2003,MotlNeitzkeATMP2003}
carries over directly, namely, 
\begin{equation}
e^{4\pi \omega}=-(1+2\cos{\pi j})\,.
\end{equation}
However, $j=2/3$ and $\cos(2\pi/3)=-1/2$, so $(1+2\cos{\pi j})=0$.  If
we add rotation to the hole, i.e., if $B\neq0$, a similar analysis
implies the same result, namely, there are no asymptotic QN
frequencies for these acoustic holes.

This result is quite hard to grasp.  It may denote either that the
real part grows without limit, or that it does not converge to a
finite value. Whatever is the correct conclusion, an application of
Hod's conjecture to these acoustic holes looks hopeless.  The reason
for this can perhaps be understood. In the analogue case, the laws of
black hole thermodynamics cannot be recreated. It was argued by Visser
\cite{VisserPRL1998} that black hole thermodynamics is directly
related to Einstein equations.  Undoubtely, a well-known defect of
analogue models is the fact that they can replicate the kinematical
atributes of general relativity, but not the dynamics disclosed by
Einstein equations.  To obtain Hod's results \cite{HodPRL1998} one has
to assume a thermodynamic relation between black hole area and
entropy. The lack of any such link for analogue models could account
for the nonexistent relation between the QNM spectrum and area
quantization.

\subsection*{4.6 Power-law tails}

After the exponential QNM decay characteristic of the ringdown
phase, black hole perturbations decay with a power-law tail
\cite{PricePRD1972} (see also \cite{CDLYPRD2004})
due to the backscattering off the background
curvature. One expects the same phenomenon holds in acoustic
holes. Thus, we have computed the late-time tails of wave propagation 
in the draining bathtub and 
found that the field falloff at
very late times is of the form 
\begin{equation}\label{dec}
\psi\sim t^{-(2m+1)}\,,
\end{equation}
where $m$ is the azimuthal angular number
\cite{BertiCardosoLemosPRD2004}. This time exponent is 
a feature
of any $(2+1)$-dimensional spacetime, and not only
of an acoustic hole
\cite{CDLYPRD2004}.

\section*{5. Superresonance in the draining bathtub acoustic hole}

Superradiance is a general phenomenon in physics.
Zel'dovich \cite{ZeldovichJETPLett1971,ZeldovichJETP1972} pointed out
that a cylinder made of absorbing material and rotating around its
axis with frequency $\Omega$ can amplify modes of scalar or
electromagnetic radiation of frequency $\omega$, provided the
condition
\be\label{suprad}
\omega<m\Omega\,,
\ee
where $m$ is the azimuthal wave number with respect to the axis of
rotation, is satisfied. This inequality is easier to 
understand if we work with  periods instead of frequencies. 
Putting $T=2\pi/\omega$ for the period of the wave, and 
$P=2\pi/\Omega$ for the period of the cylinder rotation, than the 
inequality turns into $mT>P$. This means that there is superradiance
when the wave stands in nearby a sufficient period $mT$
that it has time to absorb some of the cylinder rotational energy. 
Zel'dovich realized that, accounting for
quantum effects, the rotating object should emit spontaneously in this
superradiant regime. He further 
suggested that a Kerr black hole whose
angular velocity at the horizon is $\Omega$ will show both
amplification and spontaneous emission when the condition
(\ref{suprad}) for superradiance is satisfied. This suggestion was put
on firmer ground by a substantial body of work 
(e.g., 
\cite{PTAPJ1974,UnruhPRD1974,bekensuperr,PressTeukolskyNature1972,CDLYPRD2004}).
Black hole 
superradiance is related to the presence of an ergosphere,
allowing the extraction of rotational energy from the
black hole itself, and
it is the wave equivalent of the Penrose process for particles
\cite{PenroseNC1969}.

The possibility of finding rotational superradiance in analogues,
i.e., superresonance, was considered by Sch\"utzhold and Unruh
\cite{SchutzholdUnruhPRD2002} and Basak and Majumdar
\cite{BasakMajumdarCQG20031,BasakMajumdarCQG20032} who 
computed in the low frequency limit
the reflection coefficients 
$\omega r_H/c\ll 1$.
Here we study further the phenomenon
of superresonance for the rotating draining bathtub metric 
\cite{BertiCardosoLemosPRD2004}.

Consider an incident plane wave 
of frequency $\omega$, azimuthal wave number $m$, and
unit amplitude at infinity
coming in from infinity towards the
acoustic hole. A portion of this wave will be reflected back
by the medium to infinity, the reflection coefficient being some
complex number $R_{\omega m}$. In terms of the wave equation
(\ref{waveequation2})-(\ref{Qdef}), this implies the following
boundary condition at infinity,
\be 
H\sim R_{\omega m} e^{\ii \omega r_*}+e^{-\ii \omega r_*}
\,,\quad r\rightarrow \infty\,.
\label{asymptbeh}
\ee
At the sonic horizon ($r\to 1$, $r_*\to -\infty$) the solution 
has the following behavior
\be 
H\sim T_{\omega m} e^{-\ii (\omega-mB) r_*}\,,\quad r\rightarrow 1\,.
\label{asymptbeh1}
\ee
where $T_{\omega m}$ is the transmission coefficient. 
It gives
the percentage of 
the wave that has passed through  the horizon
into the hole. 
Note that the wave equation for this superresonance problem 
and for the QNM problem is the same. However,
each problem has their own distinct boundary conditions.

\begin{figure}
\centerline{\mbox{
\includegraphics[height=6.0cm]{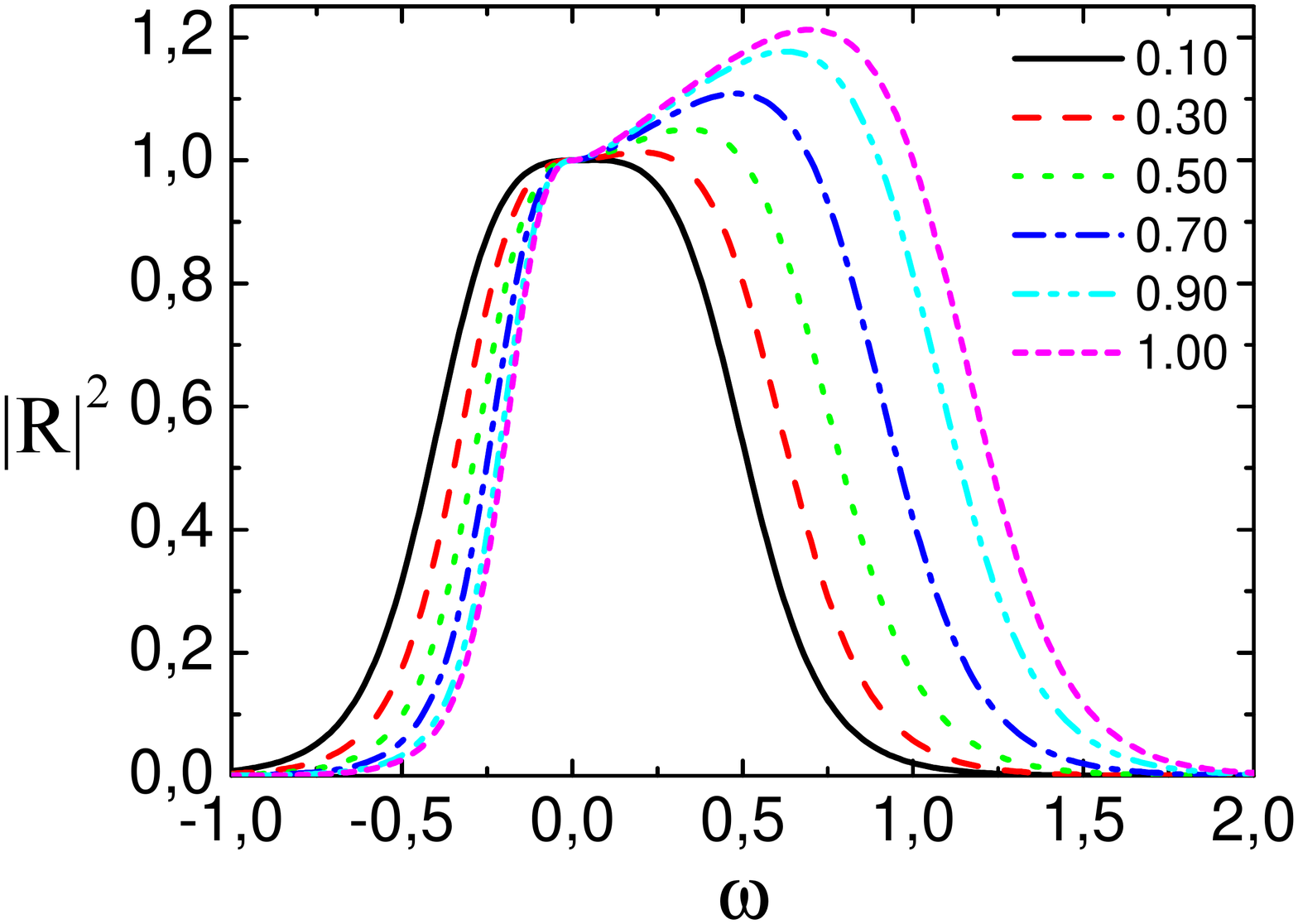}
\includegraphics[height=6.0cm]{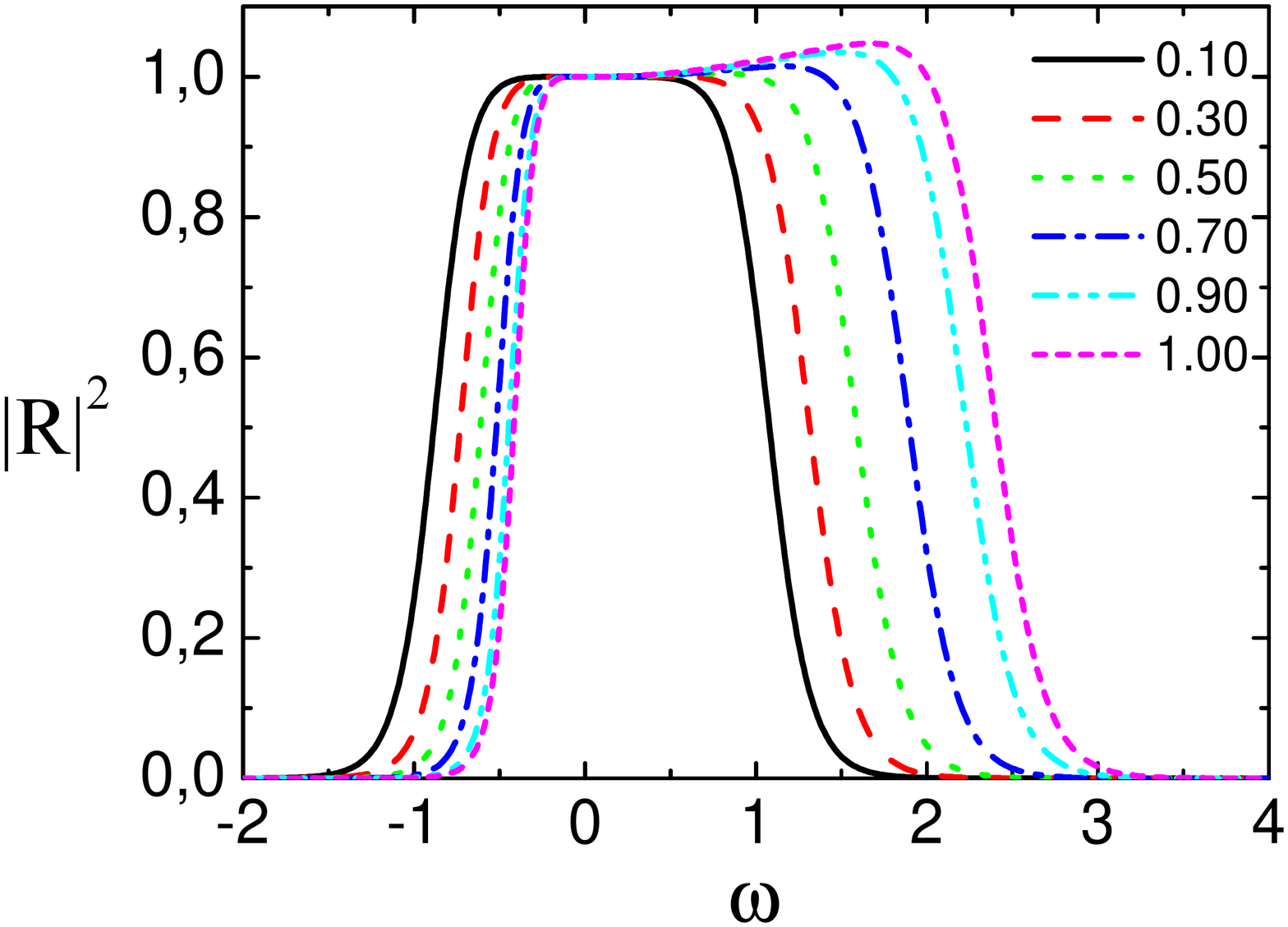}
}}
\centerline{\mbox{
\includegraphics[height=6.0cm]{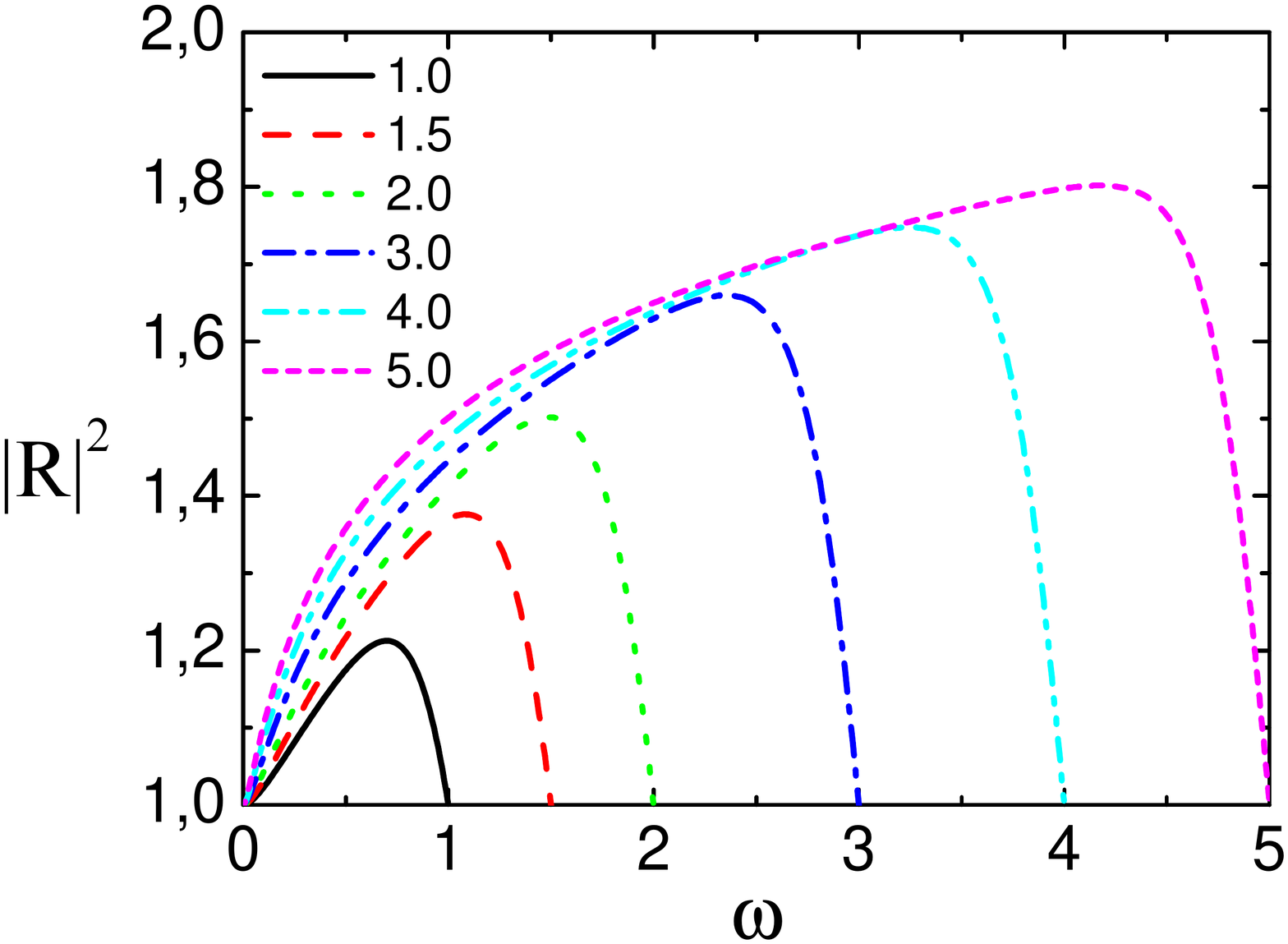}
\includegraphics[height=6.0cm]{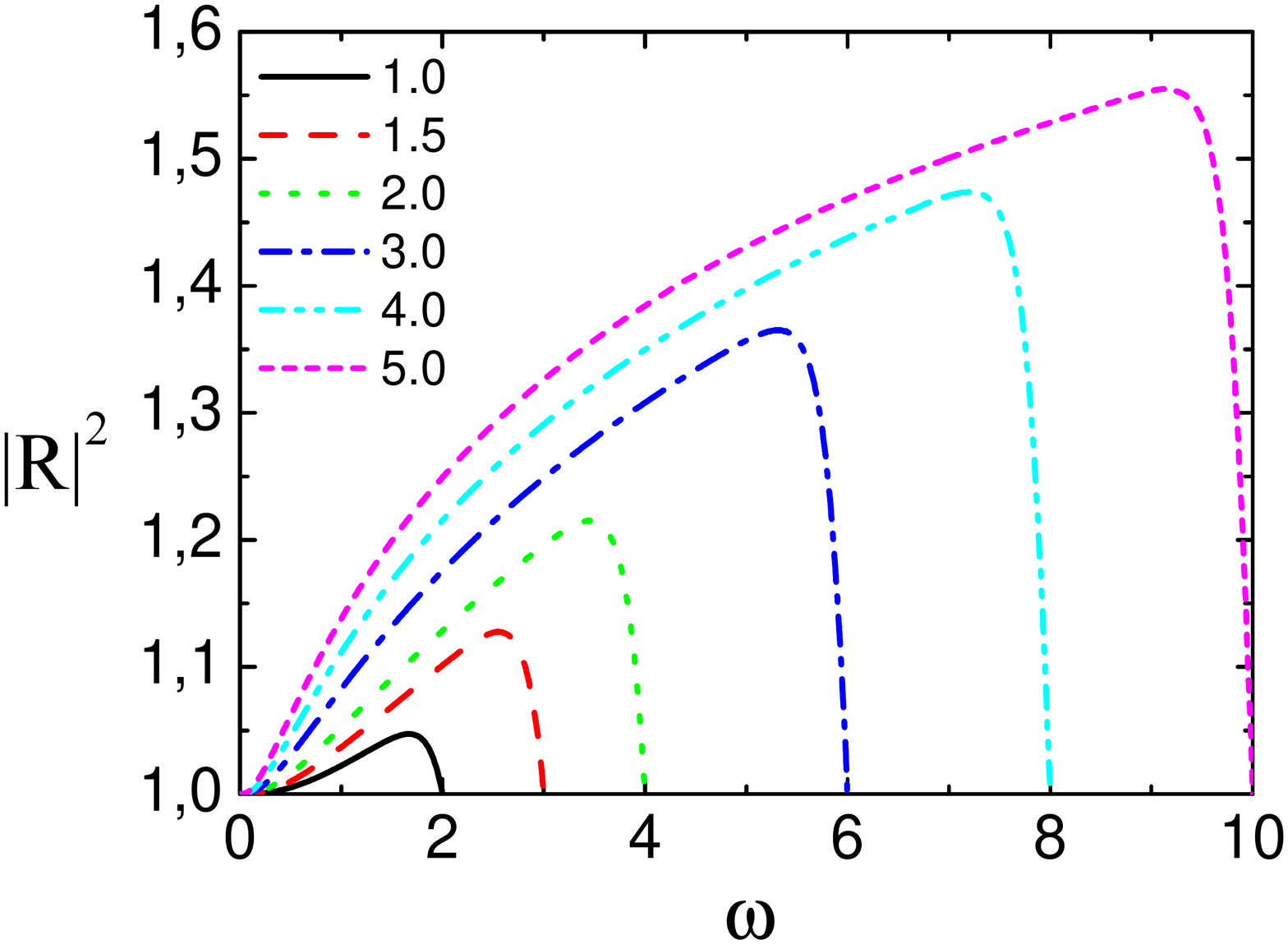}
}}
\caption{\small
Plots for the superresonant
amplification $|R_{\omega 1}|^2$
as a function of the wave frequency $\omega$ of the rotating 
draining bathtub acoustic hole. 
Top: Small rotation,
$B\leq1$, $\left|R_{\omega m}\right|^2$ as a function of
$\omega$ for $m=1$ (left plot) and $m=2$ (right plot). Each curve
corresponds to a different value of $B$. 
The reflection coefficient decays
exponentially away from $\omega_{SR}=mB$
(the critical frequency for superresonance).
For $B=1$ the maximum amplification is 21.2$\,$\%
in the $m=1$ case and 4.7$\,$\% in the $m=2$ case.
Bottom: High rotation, 
$B>1$, $\left|R_{\omega
m}\right|^2$ as a function of $\omega$ for $m=1$ (left plot) and
$m=2$ (right plot). Each curve corresponds to a different value of
$B$.  The plots show that for $B>1$ superresonant
amplification is very efficient.}
\label{superress}
\end{figure}

The Wronskian of a solution of (\ref{waveequation2})
is constant since the equation has no first derivative in 
the radial coordinate. So compute it
at the sonic horizon and at infinity and equate the computations. 
From
Eqs. (\ref{asymptbeh}) and (\ref{asymptbeh1}) 
one finds the  following condition,
\be
1-\left|R_{\omega m}\right|^2=
\left(1-\f{mB}{\omega}\right)\left|T_{\omega m}\right|^2\,,
\ee
which can be seen as an energy conservation equation.
For $\omega <mB$ the reflection coefficient obeys
$\left|R_{\omega
m}\right|^2>1$, which means there is superresonance. 
Refined 
conditions at $r_H$ and at
infinity can be used to 
compute numerically $R_{\omega m}$ \cite{BertiCardosoLemosPRD2004}.

Results for the rotating draining bathtub metric are shown in
Fig.~\ref{superress}. The plots on the left show the superresonant
amplification $|R_{\omega 1}|^2$ for $m=1$ for some $Bs$, while the
plots on the right show the superresonant amplification $|R_{\omega
1}|^2$ for $m=2$.  The plots show that the reflection coefficient
$|R_{\omega m}|^2\geq 1$ in the superresonant regime, $0<\omega<mB$,
as it should.  There are other interesting points.  As $B$ is increased
the reflection coefficient also increases.  For fixed $B$, there is a
maximum for the reflection coefficient $\left |R_{\omega m}\right|^2$
at $\omega \sim mB$, decaying exponentially afterward as a function of
$\omega$ outside the superresonant interval. This is the analogue of
the superresonant amplification behavior for the Kerr metric
\cite{PTAPJ1974}.  In particular, zooming in one finds that, for $B=1$
and $m=1$ the maximum amplification is 21.2$\,$\%, and for $B=1$ and
$m=2$ the maximum amplification is 4.7$\,$\%.  See also the work of
Oliveira, Dolan, and Crispino \cite{DolanOliveiraCrispino2011} for
other developments.

\section*{6. The sonic bomb and plant: Superresonant instabilities 
in the rotating draining  bathtub  a\-cous\-tic hole}

\subsection*{6.1 Generic features}

Kerr black holes are stable objects, but 
inside a mirror box 
they can build up an instability due to 
superradiance, originating
a black hole bomb \cite{PressTeukolskyNature1972,CDLYPRD2004}
(see also \cite{PTAPJ1974}). Indeed, a detailed analysis performed by
Cardoso, Dias, Lemos and Yoshida \cite{CDLYPRD2004} for the Kerr black
hole showed that the black hole bomb can be characterized by a set of
complex resonant frequencies, the boxed quasinormal modes (BQNMs),
which are responsible for the mode exponential growth 
and the bomb explosion.

The rotating draining  bathtub
acoustic hole also possesses an 
ergoregion which allows as well 
the possibility to make the system
unstable. Suppose the system is enclosed in a reflecting mirror 
box with 
constant radius $r_{\rm o}$.  
Now, throw into the hole a wave of frequency $\omega=
\omega_R+{\rm i}\, \omega_I$, such that
$\omega_R<mB$. The wave is amplified 
by superresonant scattering,  
depleting the
hole's rotational energy, and travels back to the mirror. There, it
is reflected and moves again into the hole, now
with increased amplitude. Through iterated reflections, the waves'
amplitude  grows exponentially with time. 
This is the analogue of the black hole bomb, it is the 
sonic bomb. Of course, the bomb can turn into a power plant
if good use is made of the stored amplified energy.

One can argue in slightly more detail.  The condition for the
existence of standing waves in the region enclosed by the mirror is
that the real part of a BQNM is proportional to $1/r_{\rm o}$, i.e.,
$\omega_R\sim1/r_{\rm o}$, where $r_{\rm o}$ is the mirror
radius. More precisely, defining $\lambda$ as
the wavelength of the standing wave, the condition is 
$\lambda \buildrel<\over\sim r_{\rm o}$, or
$\omega_R\buildrel>\over\sim 1/r_{\rm o}$, which in turn gives $r_{\rm
o}\buildrel>\over\sim 1/\omega_R$.  Now, the imaginary part of a BQNMs
is proportional to $(\omega_R-mB)$, and so in order for the system to
become unstable the system must be within the superresonant condition,
$\omega_R<mB$. Compounded with the standing wave condition this
implies that the mirror should be located at a radius $r_{\rm
o}\gtrsim 1/mB$ in order for the system to become unstable. As time
passes rotational energy is extracted from the system, the hole spins
down and $B$ decreases.  For any given $r_{\rm o}$ the instability is
eventually blocked, since the condition $r_{\rm o}\gtrsim 1/mB$ is no
longer obeyed.

\subsection*{6.2 Equation and boundary conditions}

To solve the problem mathematically one has to solve once more the
wave equation (\ref{waveequation2})-(\ref{Qdef}), but again with
different boundary conditions. Assume that one is in the presence of a
perfectly reflecting mirror, and impose two boundary
conditions. At $r_{\rm o}$ one imposes a zero field, 
and at the horizon one demands the presence of purely ingoing
waves, i.e., 
\be
H(\omega_{BQN},r_{\rm o})=0\,, \quad
H(r_H)\sim e^{-\ii(\omega_{BQN}-Bm)r_*}\,, 
\ee
where  $\omega_{BQN}$ are the complex BQNM frequencies.
Then one can integrate  the
wave equation (\ref{waveequation2})-(\ref{Qdef}) 
outwards
from the horizon. One can do it analytically and 
numerically.

\subsection*{6.3 Results}

\subsubsection*{6.3.1 Analytical results}

A matched asymptotic expansion
is used to calculate
analytically the unstable modes of the
scalar field.  One assumes
$1/\omega\gg r_H$, i.e., the wavelength of the scalar field is much larger
than the typical size of the 
acoustic hole. One considers the near region $r-r_H\ll
1/\omega$ and the far region $r-r_H\gg r_H$, and
solves the radial equation
for the two regions  matching the solution in the overlapping region,
$r_H\ll r-r_H\ll 1/\omega$.

In the near region the solution for $H$ is some function 
with two terms, one term
growing with $r$ and the other decreasing with $r$. 
In addition, there is one
constant of integration
(one boundary condition has been used). In the
far region the solution for $H$ is a linear combination of Bessel
functions with two constants, and again one term growing with $r$ and
the other decreasing with $r$. Matching the two solutions term by term
gives a relation between the constants, in particular between the two
constants of the far region solution as a function of the hole
parameters and $\omega$. Imposing further the box mirror condition for
the far-field solution one gets another relation between the two
constants of the far region solution as a function of the hole
parameters, $\omega$, and $r_{\rm o}$.  Thus a relation between $\omega$
and $r_{\rm o}$ (and the hole parameters) is obtained.
The result for the rotating draining bathtub is
\beq
\omega_{BQN}=\omega_R+\ii\omega_I,
\eeq
where 
\beq
\omega_R=\frac{j_{m,n}}{r_{\rm o}}\,,\quad
\label{realom}
\eeq
\beq
\omega_I=-C\,\left(\omega_R-mB\right)\,,
\label{imom}
\eeq
the $j_{m,n}$'s are zeros of the Bessel function of 
integer order $m$ 
(see Table \ref{tab:bessel}), with $n=0$
being the fundamental mode, and $C=C(m,r_H,r_{\rm o})$ is 
some complicated function of no 
importance here. 
For a Kerr black hole bomb one gets
$\omega_R=j_{l+1/2,n}/r_{\rm o}$
\cite{CDLYPRD2004}, instead of (\ref{realom}).

Now, the scalar field $\psi$ has time dependence
${\rm e}^{-\ii\omega t}$, i.e., 
\beq
\psi\sim H\,{e}^{-\ii\omega_R t}\,
{e}^{\omega_I t}\,.
\eeq
Then  
$\omega_I>0$ means, from Eq.~(\ref{imom}),
$\omega_R<mB$, the amplitude of the field
grows exponentially and the BQNM becomes unstable
with a growing time $\tau\sim 1/\omega_I$. On the other hand, 
$\omega_I>0$ implies $\omega_R>mB$, and the amplitude 
decays.

Since $\omega_R=\frac{j_{m,n}}{r_{\rm o}}$, the
wave frequency is proportional to $1/r_{\rm o}$.
If the distance at which the mirror is located decreases,
the allowed wave frequency increases, and there will thus be a
a critical $r_{\rm o}$, $r_{\rm oc}$ say, 
at which the BQN frequency no longer satisfies the
superresonant condition.
Thus, with the superresonance condition
$\omega_{BQN}<mB$, the instability switches
off at 
\be
r_{\rm oc}\simeq \frac{j_{m,n}}{mB}\,. 
\label{bomboff}
\ee
Since $j_{m,n}$ increases linearly with $m$ 
(which is correct for high
$m$'s, otherwise see Table \ref{tab:bessel} for low $m$'s) 
this means that the critical radius is practically 
$m$-independent. This is about 
what one can do analytically.

\begin{table}
\centering
\vskip 12pt
\begin{tabular}{@{}c|c|c|c@{}}  
\hline
\hline
$m$ &$j_{m,0}$ &$j_{m,1}$ &$j_{m,2}$\\
\hline
\hline
1 &3.83171 &7.01559  &10.17347\\
2 &5.13562 &8.41724  &11.61984\\
3 &6.38016 &9.76102  &13.01520\\
4 &7.58834 &11.06471 &14.37254\\
5 &8.77148 &12.33860 &15.70017\\
\hline 
\hline
\end{tabular}
\caption{\small
The first few values of the zeros of the Bessel function of 
integer order $m$, $j_{m,n}$, are given.}
\label{tab:bessel} 
\end{table}

\subsubsection*{6.3.2 Numerical results}

\begin{figure}
\centerline{\mbox{
\includegraphics[height=5.2cm]{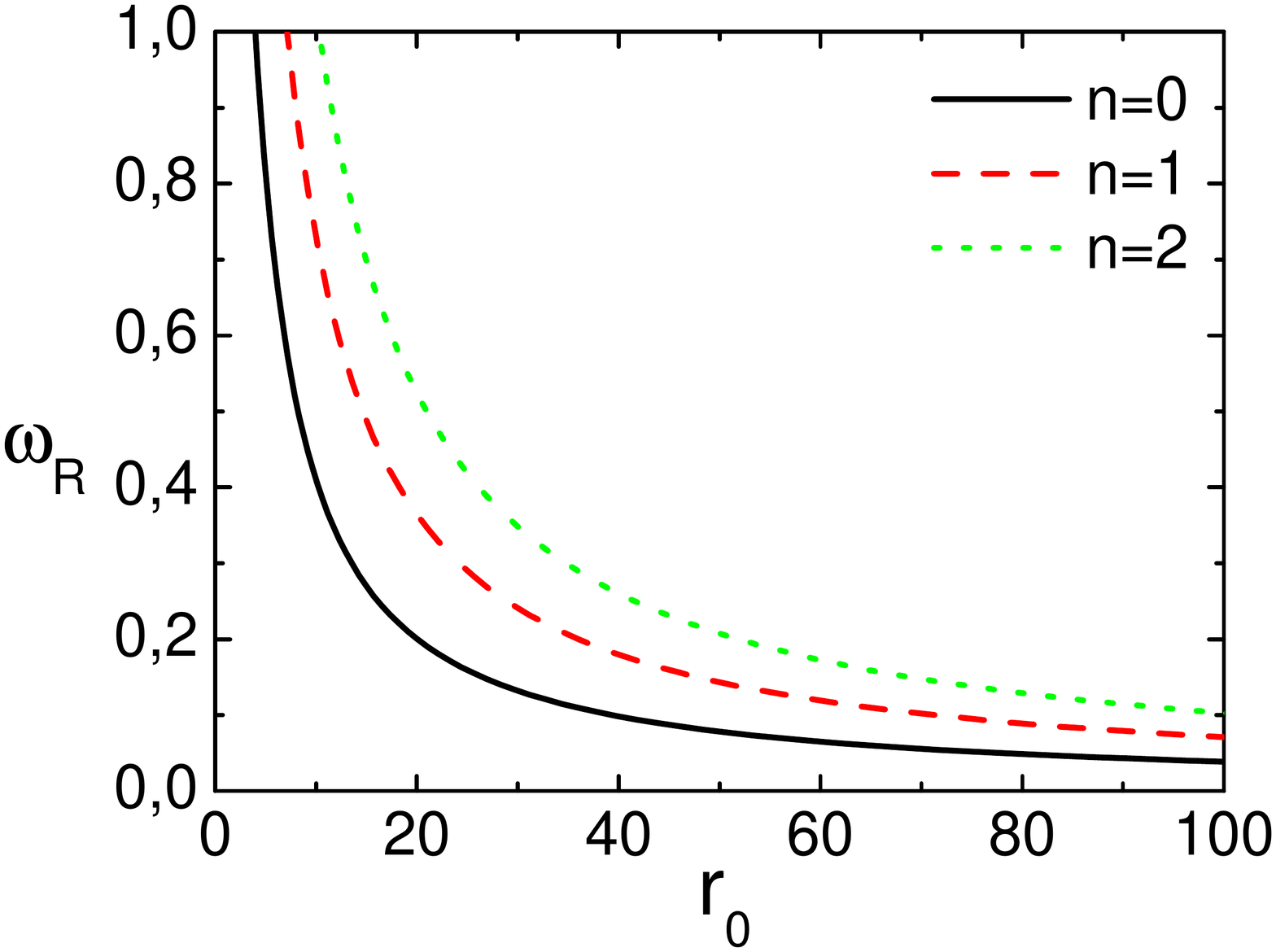}
\includegraphics[height=5.2cm]{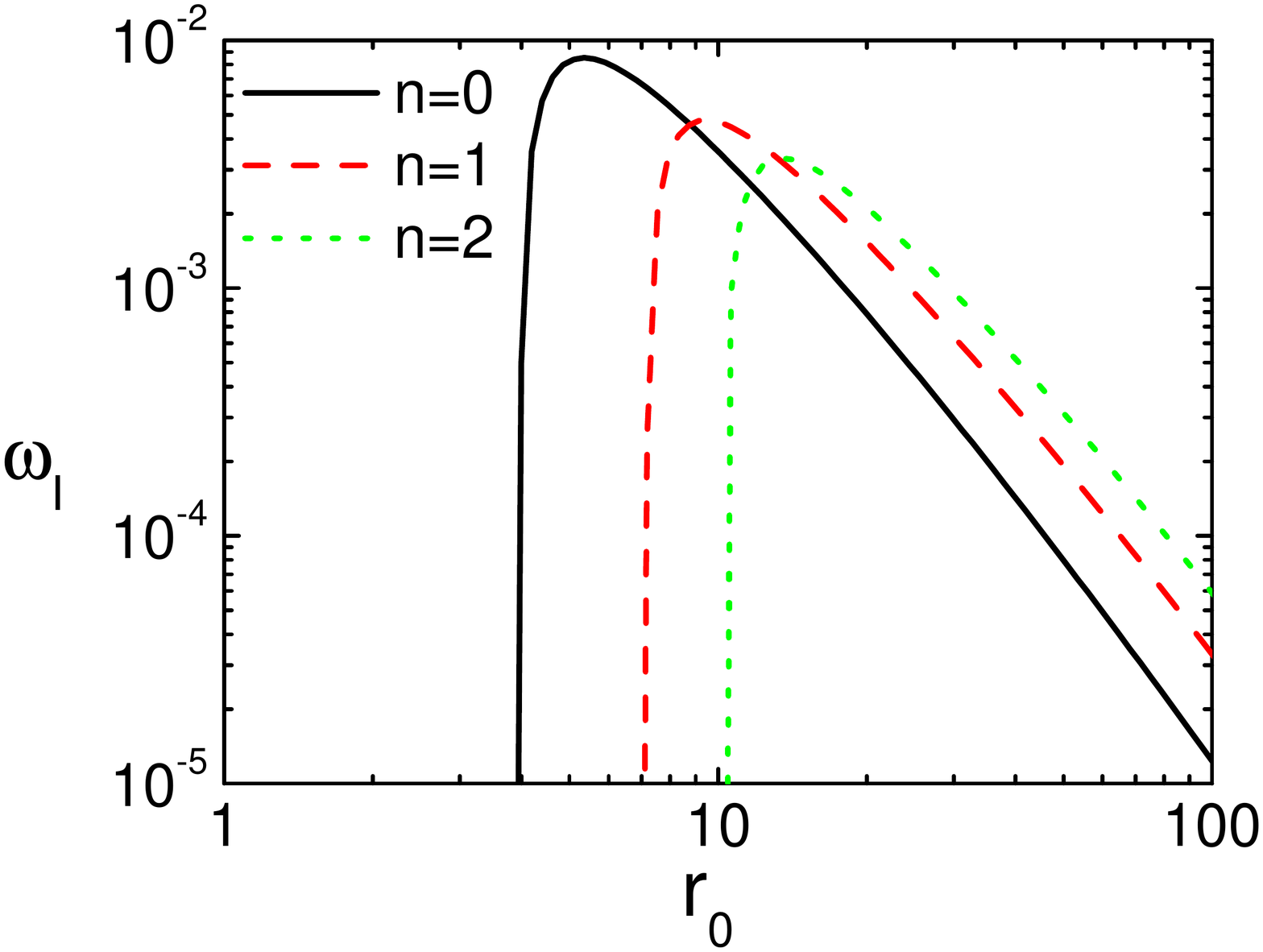}
}}
\caption{\small The   
frequencies of the BQNMs for $B=1$ and
$m=1$  as a function of
mirror location $r_{\rm o}$ for the rotating 
draining bathtub acoustic hole.
The left plot shows the real frequencies
and the right plot 
the imaginary ones.
The three curves presented are 
for the fundamental mode $n=0$, 
and the first and second overtones
$n=1,2$, respectively.
One sees that $\omega_R$ scales
as $1/r_{\rm o}$, as predicted analytically. 
The imaginary frequency $\omega_I$ crosses
zero, and the instability abruptly evanesces at 
the critical radius $r_{\rm oc}$.}
\label{BQNM1}
\end{figure}

Numerically, one can perform calculations of the unstable modes of the
scalar field by using, e.g, a Runge-Kutta method. The numerical
results are in perfect concordance with the analytical
predictions. In particular, this is true 
for small rotational parameters $B$. Note, however, 
that 
the the real part of the BQN frequencies has a  
weak $B$-dependence.
We picked  $B=1$, and $m=1$, as an
emblematic case, see
Fig.~\ref{BQNM1}.  Note that 
when $\omega_I$ crosses zero the instability
suddenly disappears. This happens
at the critical radius $r_{\rm oc}$, anticipated analytically.
From the right plot of 
Fig.~\ref{BQNM1}
we see that for any BQNM, the instability becomes more and
more efficient as the mirror radius $r_{\rm o}$ becomes smaller, until
eventually the mirror radius becomes small enough that the instability
evanesces.
\begin{figure}
\centerline{\mbox{
\includegraphics[height=5.2cm]{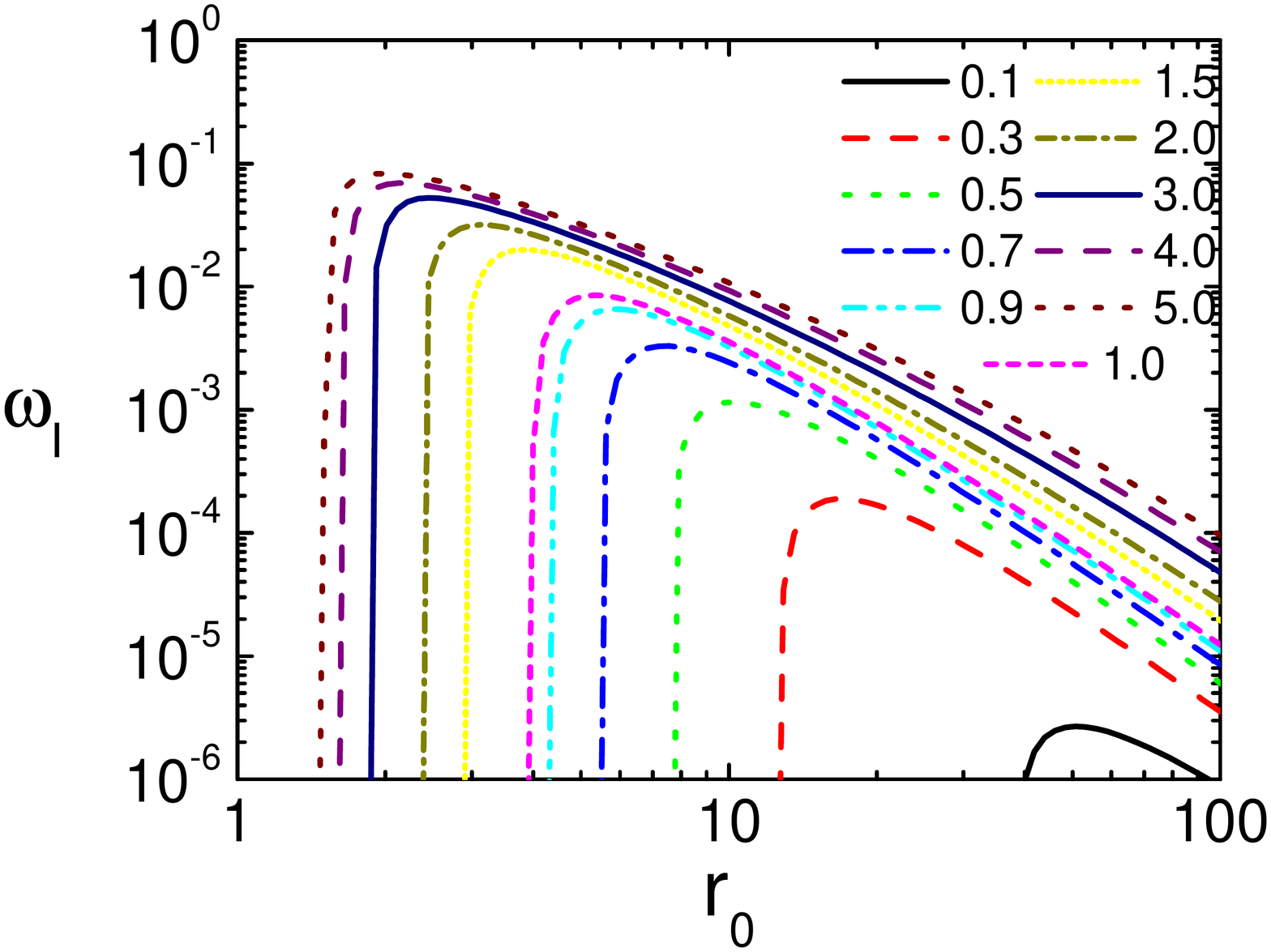}
\includegraphics[height=5.2cm]{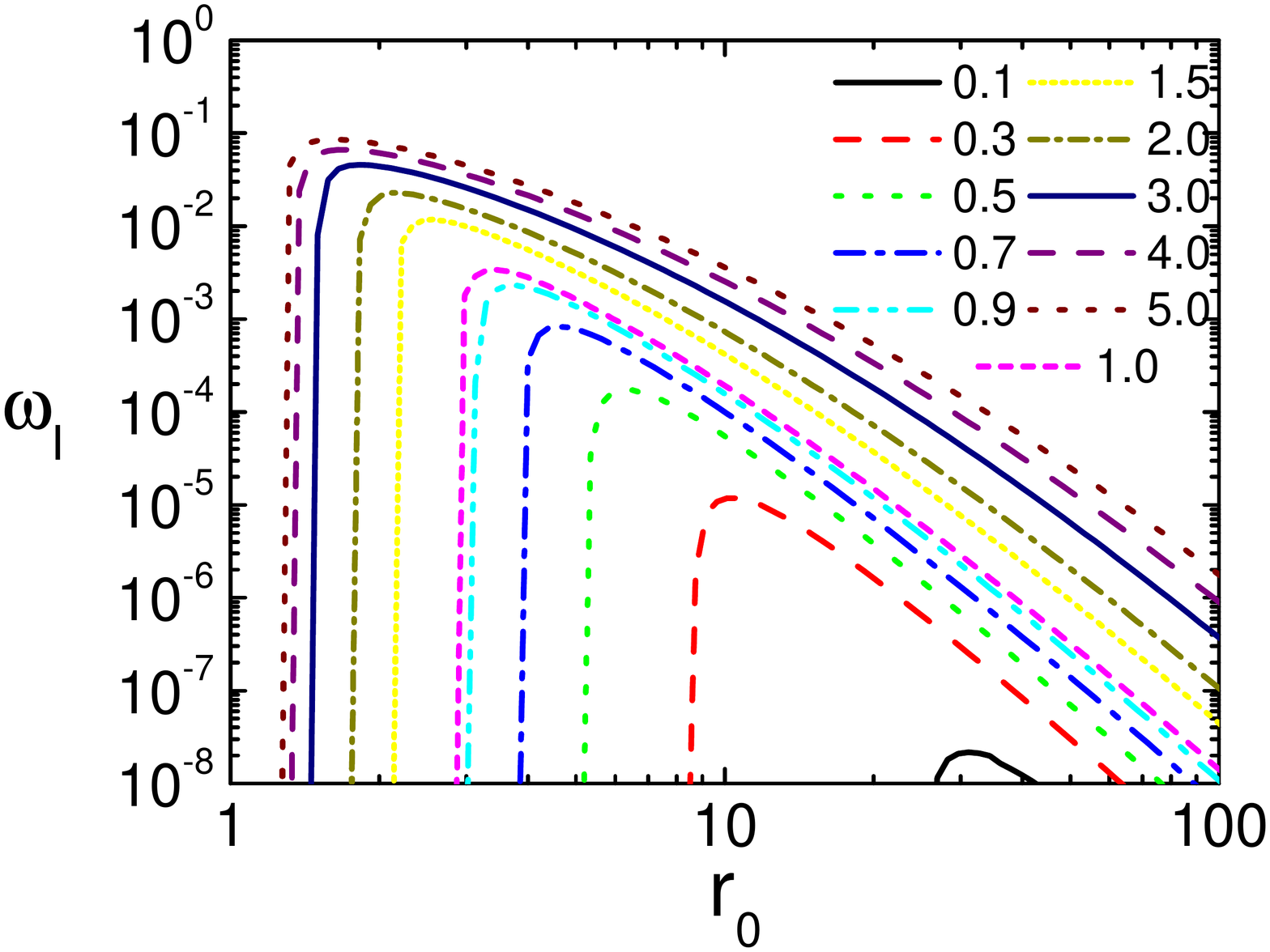}
}}
\caption{\small Imaginary frequencies $\omega_I$
of the fundamental mode $n=0$ for
BQNMs as a function of 
mirror location $r_{\rm o}$ for different values of
the rotation parameter $B$ of the rotating 
draining bathtub acoustic hole. The left plot is for $m=1$, the
right plot for $m=2$. Note the difference of scale
for the $\omega_I$
axis in each case. The imaginary frequency $\omega_I$ 
gives then the corresponding growth timescale.}
\label{BQNM2}
\end{figure}

\subsubsection*{6.3.3 Remarks and orders of magnitude calculations}

Some comments should be made:

(i) The real part of the BQN frequency grows, albeit slowly, with
growing overtone number $n$. This means that higher overtones become
stable at larger distances. In addition, they earn a smaller maximum
growing rate.

(ii) With notable precision, the instability terminates when the
critical radius is achieved, as concluded by the analytical formula
complemented by the superresonance condition, that is: $r_{\rm
oc}\simeq \frac{j_{m,n}}{mB}$.  We have also found from the analytical
formula that since $j_{m,n}$ increases linearly with $m$ this means
that the critical radius is almost $m$-independent. This is attested
numerically.  However, at least when $B\lesssim 1$, the instability is
not so proficient for high $m$ as it is for small $m$. This is made
clear in Fig.~\ref{BQNM2}, where the $\omega_I$, and thus
the growth rates, of the $m=1$ and
$m=2$ fundamental $n=0$ modes for rotating  acoustic holes are shown.
As an example take $B=0.3$. Then the maximum growth rate for $m=1$ is
$\sim 10^{-4}$, and for $m=2$ is $\sim 10^{-5}$.  When $B>1$, the
growth rate for $m=2$ becomes in effect of the same order as the
growth rate for $m=1$, as can be verified from the curves
corresponding to $B=5.0$.

(iii) One can find the response of the process by performing a
straightforward calculation.  
When the instability arises the acoustic hole is
releasing both energy and angular momentum, and the equation that
connects these two quantities is $\Delta E\sim B \Delta J$.  As $B$ is
decreased the critical radius $r_{\rm oc}$ increases.  One now
benefits from the mechanism itself by choosing a large $B$ and an
initial radius $r_{\rm o}$ for the mirror such that $r_{\rm o}\simeq 2
r_{\rm o{\rm max}}$, where $r_{\rm o{\rm max}}$ is defined as the
radius of maximum growth timescale at the given $B$.  As the hole
expels angular momentum $\Delta J$, $B$ is reduced. Given a mirror
position $r_{\rm o}$, the bomb finally shuts up when 
the condition $r_{\rm
oc}\simeq \frac{j_{m,n}}{mB}$ is obeyed. Measuring the difference in
initial and final angular momentum $\Delta J$ the extracted energy
$\Delta E$ can be ascertained.

To have a hint of the orders of magnitude that can 
appear in this experiment, let us adopt $m=1$ and an
acoustic hole rotation parameter $\widehat{B}=1$, where we have
recovered the hats, where hats distinguish dimensionless quantities
from quantities with physical units.  Let us then examine
some 
characteristic parameters for the gravity wave analogue presented in
\cite{SchutzholdUnruhPRD2002}. So put, $r_H\sim 1\,$m and $c\sim
0.31\,$m~s$^{-1}$.  Then $B\sim 0.31\,$m$^2$s$^{-1}$.  Once again, to
full use the superresonance backscattering process the mirror should
be positioned close to the maximum of $\omega_I$, but not precisely at
it. For instance, if the mirror is located at $\widehat r_{\rm
o}\simeq 10\sim 10\,$m (cf. the right plot of Fig.~\ref{BQNM2}), it is
obtained $\widehat \omega_I\simeq 4\cdot 10^{-3}$ (in this case the
maximum $\widehat\omega_I$ would be $\widehat \omega_I=8.5 \cdot
10^{-3}$). This corresponds to a growth time $\tau\simeq 800\,{\rm
s}\simeq 13\,{\rm minutes}$ (at the maximum $\tau\simeq 6$~minutes is
obtained). In rough terms, this indicates that the amplitude doubles
every 13 minutes, and will be magnified by several orders of magnitude
on timescales of a few hours.  To shorten the characteristic timescale
to seconds, one has to tune the horizon radius and wave velocity.  For
instance, working with $r_H \sim 0.1\,$m and $c\sim
10\,$m$\,$s$^{-1}$, a typical timescale of about 2 seconds is
found. This seems an acceptable timescale to discern a sonic hole bomb
in the lab.

(iv) 
As a final comment, we draw the attention to the fact
that when we describe the acoustic hole
surrounding by
a reflecting mirror, we are thinking of 
a universal kind of mirror.
In the shallow basin gravity wave analogue
\cite{SchutzholdUnruhPRD2002}, the circular mirror 
reflects gravity water waves and 
is supposed to be a rubber
band of radius $r_{\rm o}$.  A different 
option to realize a mirror could incorporate
modifications in the sound waves
dispersion relation, as would happen 
in the case of a change of
the height of the basin \cite{SchutzholdUnruhPRD2002}.
Thus, given the right environment,
one can build up an equipment that
accumulates energy in the manner
of amplified sound waves.  This can be put
to bad use as in a bomb, or to good use as in a sonic plant.

\section*{7. Conclusions}

We have considered the
$(2+1)$-dimensional draining bathtub metric, 
a rotating acoustic hole metric, the analogue to the 
Kerr metric, and studied its QNMs,  
its superresonance features, its instabilities 
when surrounding it by a mirror and the possibility 
of turning it into a sonic bomb or a power plant. 
By adding a trivial third dimension, the setup can be 
transformed into a $(3+1)$-dimensional draining bathtub.
There are other metrics that describe acoustic holes. For instance
there is the Schwarzschild analogue, a $(3+1)$-dimensional spherically
symmetric metric, called the canonical acoustic metric
\cite{VisserCQG1998}. We have also studied the 
QNMs of this metric \cite{BertiCardosoLemosPRD2004}.

\section*{Acknowledgments} We thank Luis Carlos Crispino for the
invitation and hospitality in the  II Amazonian Symposium on
Physics - Analogue Models of Gravity: 30 Years Celebration of the
publication of Bill Unruh's paper "Experimental black-hole
evaporation?", held in Universidade Federal do Par\'a, Bel\'em,
Brazil.  This work was partially funded by Funda\c c\~ao para a
Ci\^encia e Tecnologia (FCT) - Portugal, through Projects
PTDC/FIS/098962/2008 and PEst-OE/FIS/UI0099/2011.

\end{document}